\documentclass[prl,aps,showpacs,twocolumn]{revtex4}
\usepackage{epsfig}
\usepackage{bm}

\begin{document}

\title{Isotherms clustering in cosmic microwave background }

\author{  A. Bershadskii}
\affiliation{ ICAR, P.O.\ Box 31155, Jerusalem 91000, Israel}

\begin{abstract}
Isotherms clustering in cosmic microwave background (CMB) has been 
studied using the 3-year WMAP data on cosmic microwave background 
radiation. It is shown that the isotherms clustering could be produced 
by  the baryon-photon fluid turbulence in the last scattering surface.   
The Taylor-microscale Reynolds number of the turbulence is estimated 
directly from the CMB data as $Re_{\lambda} \sim 10^2$. 
\end{abstract}

\maketitle

\section{Introduction}

Luminous matter in the observable universe is strongly clustered (see, for a review
\cite{p1}). Main reasons for this apparent clustering are gravitational 
forces and an initial non-uniformity of the baryon-photon matter just before the 
recombination time. An information about this non-uniformity we can infer from the Cosmic Microwave 
Background (CMB) radiation maps. Although these maps indicate nearly Gaussian distribution of 
the CMB fluctuations, certain clustering 
may be observed already in these maps even after they have monopole and dipole set to zero. 
The clustering of the isotherms in the last scattering surface then can be used as an initial 
condition for the gravitational simulations.
One of the purposes of this paper is to describe the 
clustering quantitatively. Another related purpose is to find out underlying primordial 
physical processes, which may cause the isotherms clustering in the CMB maps. It is known 
\cite{sbar},\cite{dky} that rotational velocity perturbations in the primordial baryon-photon 
fluid can produce angular scale anisotropies in CMB radiation through the Doppler 
effect. The conclusions are relevant to arcminute scales (for large scales see \cite{jaf}). 
On the other hand, it is recently shown in \cite{sb1} that clustering is an 
intrinsic property of the turbulent (rotational) velocity field. Therefore, one should check 
whether the CMB isotherms clustering can be related to turbulent motion of the 
baryon-photon fluid just before recombination time (an observational indication of turbulent 
motion in the  baryon-photon fluid just before recombination time is given in \cite{bs1}, see also 
\cite{gibson}-\cite{kah}). 
Moreover, we will estimate the Taylor-microscale Reynolds 
number ($Re_{\lambda}$, see Appendix A) of this motion, that is $Re_{\lambda} \sim 10^2$ (let us 
recall that critical value is $Re_{\lambda} \simeq 40$ ). 
As far as we know this is first estimate of the Reynolds number obtained directly from 
a CMB-map (we have used a cleaned 3-year WMAP \cite{tag}).   \\

To find physical (dynamical) origin of the isotherms clustering in the CMB we will 
compare certain statistical properties of the CMB maps (and their Gaussian simulations) with 
corresponding properties of the fluid turbulence. It is well known that 
energetically the CMB fluctuations are dominated by acoustic perturbations of the primordial 
velocity field. The velocity field can be represented by following way
$$
{\bf u} = {\bf u_a} + {\bf u_r}
$$
where the acoustic component is a potential one ${\bf u_a} = \nabla \varphi $ and ${\bf u_r}$ is 
rotational component of the primordial velocity field. To exclude the acoustic component one can 
take operation $curl ~{\bf u}= curl ~{\bf u_r}$, that is vorticity. Average energy dissipation in turbulent 
velocity field is determined just by its rotational component \cite{my}
$$
\langle \varepsilon \rangle \sim \nu \langle (curl~ {\bf u})^2 \rangle  
$$
where $\langle ... \rangle$ means statistical average and $\nu$ is kinematic viscosity. 
Since it is expected that high frequency events in the velocity field should 
provide the most significant contribution to the turbulent dissipation and, especially, 
to its high order moments one can also expect that 
clustering in turbulent velocity field should be intimately related to fluctuations of the 
energy dissipation (so-called {\it intermittency} phenomenon \cite{my},\cite{sa}). 
Taking into account that turbulent  
velocity is nearly Gaussian \cite{my} (see also Appendix B) the clustering 
phenomenon in the turbulent velocity field should be also a Gaussian one. On the other hand, 
the intermittency phenomenon in fluid turbulence, usually associated 
with fine non-Gaussian properties of the velocity field \cite{my},\cite{sa}. 
Thus, simultaneous consideration of the cluster and intermittency characteristics can 
provide an additional valuable information about physical origin of the isotherms clustering 
in the CMB maps and about the fine non-Gaussianity of the maps. Therefore, the paper starts with an 
introduction to this subject.

\section{Clustering and intermittency}
For turbulent flows the so-called Taylor hypothesis is generally used to interpret 
the turbulent data \cite{my},\cite{sa},\cite{b}. 
This hypothesis states that the intrinsic time dependence of the wavefields 
can be ignored when the turbulence is convected past the probes at nearly constant 
speed. With this hypothesis, the temporal dynamics should reflect the spatial one, 
i.e. the fluctuating velocity field measured by a given probe 
as a function of time, $u(t)$ is the same as the velocity $u(x/\langle u  \rangle)$ 
where $\langle u \rangle$ is the mean velocity and $x$ is the distance to a position 
"upstream" where the velocity is measured at $t=0$. 
%%%%%%% FIGURE 1 %%%%%%%%%%%%%%%%%%
\begin{figure} \vspace{-0.5cm}\centering
\epsfig{width=.45\textwidth,file=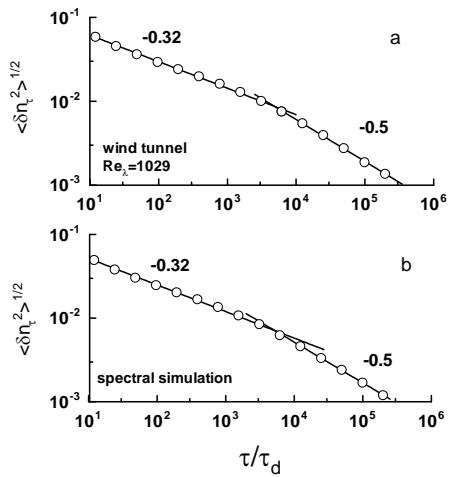} \vspace{-3.5cm}
\caption{Standard deviation of the running density fluctuations
against $\tau$ for for the velocity fluctuations measured
in the wind-tunnel (Fig. 1a) and for their Gaussian spectral simulation
(Fig. 1b).}
\end{figure}
%%%%%%%%%%%%%%%%%%%%%%%%%%%%%%%%%%%
%%%%%%% FIGURE 2 %%%%%%%%%%%%%%%%%%
\begin{figure} \vspace{-0.5cm}\centering
\epsfig{width=.45\textwidth,file=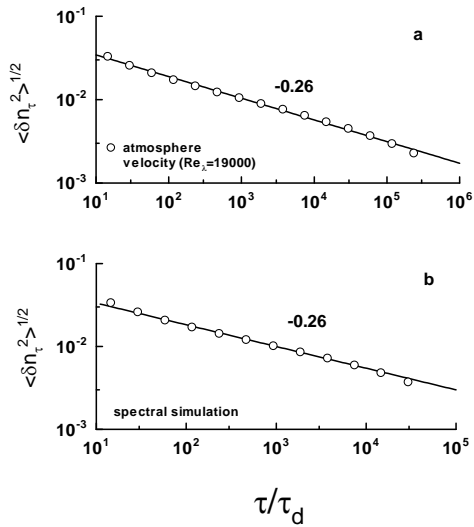} \vspace{-3.5cm}
\caption{As in Fig. 1 but for atmospheric surface layer at 
$Re_{\lambda}=19000$ (Fig. 2a) and for its Gaussian spectral simulation (Fig. 2b).}
\end{figure}
%%%%%%%%%%%%%%%%%%%%%%%%%%%%%%%%%%%

Let us count the number of 'zero'-crossing points of the signal in a
time interval $\tau$ and consider their running density
$n_{\tau}$. Let us denote fluctuations of the running density as
$\delta n_{\tau} = n_{\tau} - \langle n_{\tau} \rangle$, where the
brackets mean the average over long times. We are interested in
scaling variation of the standard deviation of the running density
fluctuations $\langle \delta n_{\tau}^2 \rangle^{1/2}$ with
$\tau$
$$
\langle \delta n_{\tau}^2 \rangle^{1/2} \sim \tau^{-\alpha_n}
\eqno{(1)}
$$
For white noise it can be derived analytically \cite{molchan},\cite{lg} that 
$\alpha_n = 1/2$ (see also \cite{sb1}).  

In Figure 1a and 2a we show calculations of the standard deviation for a turbulent 
velocity signal obtained in the wind-tunnel experiment \cite{pkw} 
and for a velocity signal obtained in an atmospheric experiment \cite{sd}) respectively.
The straight lines are drawn in the figures to indicate scaling
(1). One can see two scaling intervals in Fig. 1a. The left
scaling interval covers both dissipative and inertial ranges,
while the right scaling interval covers scales larger then the
integral scale of the flow. While the right scaling interval is
rather trivial (with $\alpha_n =1/2$, i.e. without clustering), the
scaling in the left interval (with $\alpha_n < 1/2$) indicates
clustering of the high frequency fluctuations. The
cluster-exponent $\alpha_n$ decreases with increase of
$Re_{\lambda}$, that means increasing of the clustering (as it
was expected from qualitative observations). 

To describe intermittency the standard deviation of fluctuations of the energy 
dissipation is used \cite{po} (see also a discussion in \cite{lp})
$$
\langle \delta \varepsilon_{\tau}^2 \rangle^{1/2} \sim
\tau^{-\alpha_i}  \eqno{(2)}
$$
where
$$
\varepsilon_{\tau} = \frac{\int_0^{\tau} \varepsilon (t) dt}{\tau}, \eqno{(3a)}
$$
or in the terms of the {\it discrete} time series \cite{sa}
$$
\varepsilon_{\tau}\sim \sum_{j=1}^\tau \Delta u_j^2 /\tau  \eqno{(3b)}
$$
where $\Delta u_j =u_{j+1}-u_j$. 

%%%%%%% FIGURE 3 %%%%%%%%%%%%%%%%%%
\begin{figure} \vspace{-0.5cm}\centering
\epsfig{width=.45\textwidth,file=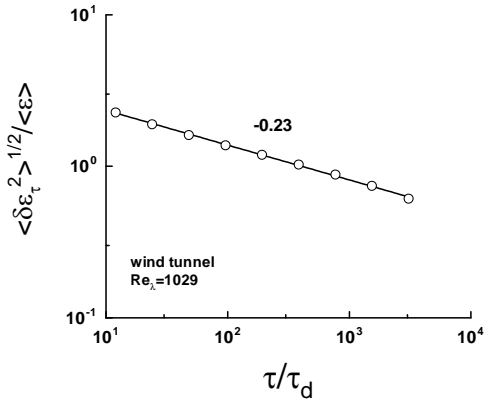} \vspace{-4.5cm}
\caption{Intermittency exponent as slope of the
straight line for the longitudinal velocity signal obtained in
the wind tunnel experiment \cite{pkw} at $Re_{\lambda} = 1029$.}
\end{figure}
%%%%%%%%%%%%%%%%%%%%%%%%%%%%%%%%%%%}
Figure 3 shows an example of the scaling (2) providing the intermittency exponent 
$\alpha_i$ for the wind-tunnel data (see \cite{po} for more details).
%%%%%%% FIGURE 4 %%%%%%%%%%%%%%%%%%
\begin{figure} \vspace{-0.5cm}\centering
\epsfig{width=.45\textwidth,file=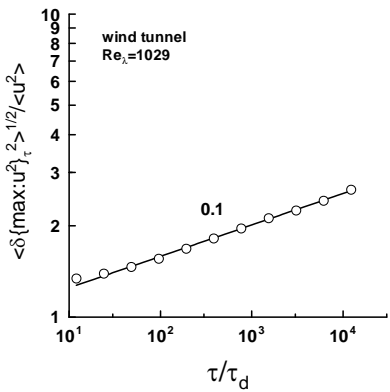} \vspace{-4cm}
\caption{ Statistical ('tail') exponent $\alpha_m \simeq 0.1$ as slope of the
straight line for the velocity signal obtained in
the wind tunnel experiment \cite{pkw} at $Re_{\lambda} = 1029$.}
\end{figure}
%%%%%%%%%%%%%%%%%%%%%%%%%%%%%%%%%%%
Events with high concentration of the zero-crossing points in the intermittent 
turbulent velocity signal provide main contribution to the $\langle \delta \varepsilon_{\tau}^2 \rangle$ 
due to high concentration of the statistically significant local maximums 
in these events \cite{bt}. Therefore, one can make use 
of statistical version of the theorem 'about mean' in order to estimate the standard deviation
$$
\langle \delta \varepsilon_{\tau}^2 \rangle^{1/2} \sim  \langle 
\delta \{\max : u^2\}_{\tau}^2 \rangle^{1/2} \langle \delta n_{\tau}^2 \rangle^{1/2}  
\eqno{(4)}
$$
where $\{\max : u^2\}_{\tau}$ is maximum of $u(t)^2$ in 
interval of length $\tau$. 
Value of $\langle \delta \{\max : u^2\}_{\tau}^2 \rangle^{1/2}$ should grow 
with $Re_{\lambda}$ (due to increasing of $\langle  u^2 \rangle$) and, 
statistically, with the length of interval 
$\tau$ (the later growth is expected to be self-similar), i.e.
$$
\langle \delta \{\max : u^2\}_{\tau}^2 \rangle^{1/2} = C(Re_{\lambda})
\cdot \tau^{\alpha_m} \eqno{(5)}
$$
where constant $C(Re_{\lambda})$ is a monotonically increasing function of $Re_{\lambda}$, 
and the statistical ('tail') exponent $\alpha_m \geq 0$ is independent on $Re_{\lambda}$. Fig. 4 
shows an example of scaling (5) providing the statistical ('tail') exponent 
$\alpha_m \simeq 0.10 \pm 0.01$ for the wind-tunnel data.
Substituting (5) into (4) and taking into account (1) and (2) we can infer 
a relation between the scaling exponents
$$
\alpha_i=\alpha_n-\alpha_m   \eqno{(6)}
$$
Unlike the exponent $\alpha_m$ the exponent $\alpha_n$ 
depends on $Re_{\lambda}$. Therefore, we can learn from (6) 
that just the clustering
of high frequency velocity fluctuations ($\alpha_n$ in (6)) is responsible for  
dependence of the intermittency exponent $\alpha_i$ on the $Re_{\lambda}$.

It is naturally consider the exponent 
functions through $\ln R_{\lambda}$: $\alpha = \alpha (\ln Re_{\lambda})$
\cite{sb1},\cite{cgh},\cite{bg} (see also Appendix A). Following to the general idea
of Ref. \cite{bg} (see also \cite{sb1}) let us expend this function for large
$Re_{\lambda}$ in a power series:
$$
\alpha (\ln Re_{\lambda}) = \alpha(\infty) + \frac{a_1}{\ln
Re_{\lambda}} + \frac{a_2}{(\ln Re_{\lambda})^2} + ... \eqno{(7)}
$$
In Figure 5 we show the calculated values of $\alpha_n$ (circles) 
and $\alpha _i$ (squares)
for velocity signals against $1/\ln Re_{\lambda}$ for $200 <
Re_{\lambda} < 20000$ (for $\alpha _i$ we also added four data 
points taken from the paper \cite{po}, where the data
were obtained in a mixing layer and in an atmospheric surface
layer). The straight lines (the best fit) indicate
approximation with the two first terms of the power series
expansion
$$
\alpha_n (\ln Re_{\lambda}) \simeq 0.1 + \frac{3/2}{\ln
Re_{\lambda}},~~~~~~~~\alpha_i (\ln Re_{\lambda}) \simeq \frac{3/2}{\ln
Re_{\lambda}} \eqno{(8)}
$$
This means, in particular, that
$$
\lim_{Re_{\lambda} \rightarrow \infty}\alpha_n \simeq 0.1,~~~~~ 
\lim_{Re_{\lambda} \rightarrow \infty}\alpha_i\simeq 0 \eqno{(9)}
$$
The closeness of the constants $a_1 \simeq 3/2$ (the slopes of the straight lines in Fig. 5)
in the approximations (8) of the cluster-exponent $\alpha_n$ and of the intermittency exponent 
$\alpha_i$ confirms the relationship (6) with $\alpha_m \simeq 0.1$ (cf. Figs. 4 and 5). 
%%%%%%% FIGURE 5 %%%%%%%%%%%%%%%%%%
\begin{figure} \vspace{-1.5cm}\centering
\epsfig{width=.45\textwidth,file=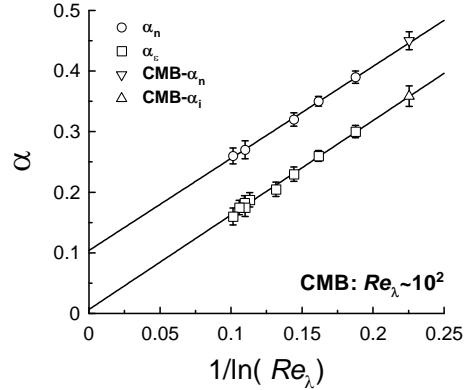} \vspace{-3.5cm}
\caption{The cluster-exponent $\alpha_n$ (circles) and intermittency exponent
$\alpha_i$ (squares) against
$1/\ln Re_{\lambda}$ for velocity signals at different values of
Reynolds number ($200 < Re_{\lambda} < 20000$). }
\end{figure}
%%%%%%%%%%%%%%%%%%%%%%%%%%%%%%%%%%%

Thus one can see from Eq. (6) that entire dependence of the intermittency exponent 
$\alpha_i$ on $Re_{\lambda}$ is determined by dependence 
of the cluster-exponent $\alpha_n$ on $Re_{\lambda}$. On the other hand, 
the cluster-exponent itself is uniquely determined 
by energy spectrum of the velocity signal at suggestion that the velocity field is 
Gaussian. However, different spectra can produce the same value of the cluster-exponent 
(in particular, the cluster-exponent is invariant to variation of the potential component 
of the velocity field, that is significant for next section).

We use spectral simulation to illustrate that energy spectrum uniquely determines 
cluster-exponent for Gaussian-like turbulent signals. In these simulations we generate Gaussian 
stochastic signal with energy spectrum given as data (the spectral data are taken 
from the original velocity signals). Figures
1b and 2b show, as examples, results of such simulation for the wind-tunnel 
velocity data (Fig. 1b) and for the atmospheric surface layer (Fig. 2b).
Namely, Figures 1a and 2a show results of calculations for the
original velocity signals and Figs. 1b, and 2b show corresponding
results obtained for the spectral simulations of the signals. This means, in particular, 
that in respect to the cluster-exponent the turbulent velocity signals exhibit Gaussian 
properties. Of course, it is not the case for the intermittency exponent 
$\alpha_i$. Namely, $\alpha_i \simeq 0.5$ for the Gaussian 
spectral simulations, independently on the spectral shape. 

\section{CMB clustering and intermittency}

The motion of the scatterers in the last scattering surface 
imprints a temperature fluctuation,
$\Delta T$, on the CMB through the Doppler effect \cite{sbar},\cite{dky}
$$
\frac{\Delta T ({\bf n})}{T} \sim \int g(L){\bf n} \cdot {\bf v_b}
({\bf x})~dL
$$
where ${\bf n}$ is the direction (the unit vector) on the sky,
${\bf v_b}$ is the velocity field of the baryons evaluated along
the line of sight, ${\bf x} = L{\bf n}$, and $g$ is the so-called
visibility. It should be noted that, in a potential flow, fluctuations
perpendicular to the line of sight lack a velocity component
parallel to the line of sight; consequently, generally there is no
Doppler effect for the potential flows. The same is not true for
rotational flows, since the waves that run perpendicular to the line
of sight have velocities parallel to the line of site \cite{ov}. 

Clustering in the turbulent velocity field in the last scattering surface, 
therefore, can produce clustering of the observed CMB isotherms. Thus (see previous 
section), the areas of strong clustering of the CMB isotherms can indicate the 
areas of the strong vorticity activity in last scattering surface. Moreover, using 
results of previous section we can, calculating corresponding cluster- and intermittency 
exponents for the CMB maps, check whether the turbulent relationship (6) is also 
valid for the CMB data, with the same value of $\alpha_m \simeq 0.1$ as for the fluid 
turbulence. In the case of positive answer we can then estimate the Taylor-microscale 
Reynolds number $Re_{\lambda}$ for the primordial turbulence in the last scattering 
surface using Fig. 5.  

We used the 3-year WMAP data cleaned from foreground contamination by Tegmark's group 
\cite{tag} (the original temperature is given in $\mu K$ units). Let us
introduce a gradient measure for the cosmic microwave radiation
temperature $T$ fluctuations that is an analogy of the turbulent energy dissipation 
measure $\varepsilon_r$ \cite{sa},\cite{b}
$$
\chi_r =\frac{\int_{v_r} (\bigtriangledown{T})^2 dv}{v_r}
\eqno{(10)},
$$
where $v_r$ is a subvolume with space-scale $r$ (this is a space analogy of Eqs. (3a,b), 
see also Introduction: $\langle \varepsilon \rangle \sim  \langle (curl~ {\bf v})^2 \rangle$). 

Then, scaling law 
$$
\langle \delta \chi_{r}^2 \rangle \sim r^{-\alpha_i}      \eqno{(11)}
$$
corresponds to the scaling law (2) and the exponent $\alpha_i$ is the CMB 
intermittency exponent. Technically, using the cosmic microwave 
pixel data map, we will calculate the cosmic microwave radiation temperature gradient
measure using summation over pixel sets (by random walks method) 
instead of integration over subvolumes $v_r$. The scaling of type (11) (if exists)
will be then written as
$$
\langle \delta \chi_s^2 \rangle^{1/2} \sim
s^{-\alpha_i}      \eqno{(12)}
$$
where the metric scale $r$ is replaced by number of the pixels,
$s$, characterizing the size of the summation set (the random walk trajectory length). 
The $\chi_s$ is a surrogate of the real 3D gradient measure $\chi_r$. It is
believed that such surrogates can reproduce quantitative
scaling properties of the 3D prototypes \cite{sa}. Analogously to the previous 
section the cluster analysis can be performed 
$$
\langle \delta n_s^2 \rangle^{1/2} \sim
s^{-\alpha_n}      \eqno{(13)}
$$

%%%%%%% FIGURE 6 %%%%%%%%%%%%%%%%%%
\begin{figure} \vspace{-0.5cm}\centering
\epsfig{width=.45\textwidth,file=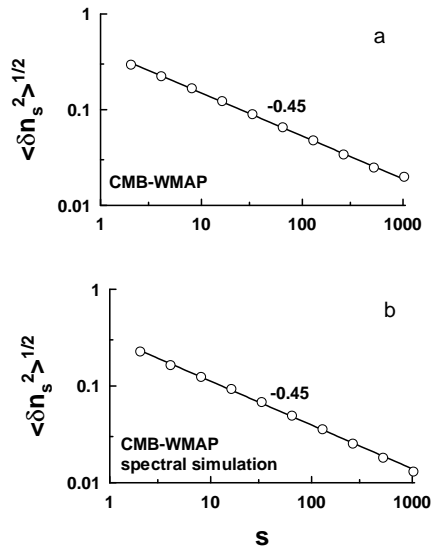} \vspace{-3 cm}
\caption{As in Figs. 1,2 but for the CMB fluctuations in the 3 year WMAP 
(Fig. 6a) and for the corresponding Gaussian spectral simulation map (Fig. 6b).}
\end{figure}
%%%%%%%%%%%%%%%%%%%%%%%%%%%%%%%%%%%
%%%%%%% FIGURE 7 %%%%%%%%%%%%%%%%%%
\begin{figure} \vspace{-0.5cm}\centering
\epsfig{width=.45\textwidth,file=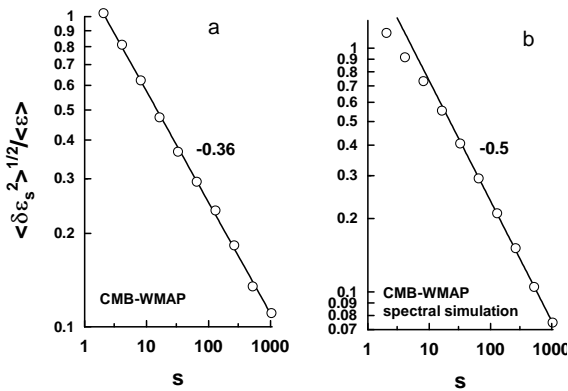} \vspace{-4.5 cm}
\caption{As in Fig. 3 but for the CMB fluctuations in the 3 year WMAP 
(Fig. 7a) and for the corresponding Gaussian spectral simulation map (Fig. 7b).}
\end{figure}
%%%%%%%%%%%%%%%%%%%%%%%%%%%%%%%%%%%

Figure 6a shows the {\it cluster}-scaling (13) for the CMB calculated for the 3-year 
WMAP map cleaned by Tegmark's group \cite{tag}. The straight line (the best fit) 
is drawn to indicate the scaling in
log-log scales. Figure 6b shows the results of analogous
calculations produced for Gaussian spectral simulation of the 3-year WMAP map. 
As it was expected the Gaussian spectral simulation produces map with the same 
cluster properties as the original map (cf. previous section). 
 
The intermittency exponent $\alpha_i$, on the contrary, takes significantly different 
values for the original CMB map and for its Gaussian spectral simulation (Figs. (7a,b)). 
One can see, that for the Gaussian map-simulation $\alpha_i \simeq 0.5 \pm 0.02$ (as it is expected, see 
previous section), while for the original CMB map $\alpha_i \simeq 0.36 \pm 0.02$ in agreement with 
the fluid turbulence case (Eq. (6) and Fig. 5). Then, using Fig. 5 we can estimate 
the Taylor-microscale Reynolds number of the primordial turbulence in the last scattering 
surface (with logarithmic accuracy) $Re_{\lambda} \sim 10^2$. 

It should be noted, that critical value (for transition from laminar to turbulent motion) is $Re_{\lambda} 
\simeq 40$ \cite{sb1}. Therefore, one can conclude that motion of the baryon-photon fluid 
at the recombination time is rather close to the critical state. This can be considered as 
an indication that the recombination process itself might be a source of the turbulence, 
which modulates the CMB. Indeed, 
the strong photon viscosity is known as the main reason for presumably laminar motion of the 
baryon-photon fluid just before recombination time. Therefore, if the recombination starts from 
appearance of the small localized transparent spots, then activity related to these spots can 
result in turbulent motion of the entire bulk of the baryon-photon fluid at the recombination time.   \\

I thank K.R. Sreenivasan for inspiring cooperation.
I also thank  D. Donzis, C.H. Gibson and participants of the workshop 
"Nonlinear cosmology: turbulence and fields"  for discussions.

\begin{center}
{\bf Appendix A}
\end{center}

Thin vortex tubes (or filaments) are the most prominent
hydrodynamical elements of turbulent flows \cite{batchelor}. 
The filaments are unstable in
3-dimensional space. In particular, a straight line-vortex can
readily develop a kink propagating along the filament with a constant speed. 
To estimate the velocity of propagation of such a kink 
let us first recall the properties of a ring vortex \cite{batchelor}. 
Its speed $v$ is related to its diameter $\lambda$ and
strength $\Gamma$ through
$$
v=\frac{\Gamma}{2\pi \lambda} \ln \left( \frac{\lambda}{2\eta}
\right),        \eqno{(A.1)}
$$
where $\eta$ is the radius of the core of the ring and
$\lambda/2\eta \gg 1$ (see figure 8).

If, for instance, a straight line-vortex develops a kink with a
radius of curvature $\lambda/2$, then self-induction generates a
velocity perpendicular to the plane of the kink. This velocity can
be also calculated using (A.1).
%%%%%%% FIGURE 8%%%%%%%%%%%%%%%%%%
\begin{figure} \vspace{-2cm}\centering
\epsfig{width=.45\textwidth,file=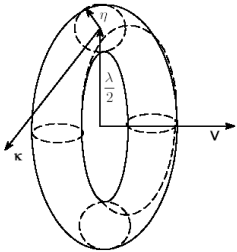} \vspace{-4.5 cm}
\caption{The vortex ring and its velocity.}
\end{figure}
%%%%%%%%%%%%%%%%%%%%%%%%%%%%%%%%%%%
One can guess that in a turbulent environment,
the most unstable mode of a vortex tube with a thin core of length
$L$ (integral scale) and radius $\eta$ (Kolmogorov or viscous scale), will be
of the order $\lambda$: Taylor-microscale \cite{my}. Then, the
characteristic scale of velocity of the mode with the space scale
$\lambda$ can be estimated with help of equation (A.1). Noting
that the Taylor-microscale Reynolds number is defined as \cite{my}
$$
Re_{\lambda}=\frac{v_0 \lambda}{\nu},   \eqno{(A.2)}
$$
where $v_0$ is the root-mean-square value of a component of
velocity. It is clear that the velocity that is more relevant (at
least for the processes related to the vortex instabilities) for
the space scale $\lambda$ is not $v_0$ but $v$ given by (A.1).
Therefore, corresponding effective Reynolds number should be
obtained by the renormalization of the characteristic velocity in
(A.2) \cite{sb1}, as 
$$
Re_{\lambda}^{eff}=\frac{v \lambda}{\nu} \sim \frac{\Gamma}{2\pi
\nu}\ln \left( \frac{\lambda}{2\eta} \right). \eqno{(A.3)}
$$

It can be readily shown from the definition that
$$
\frac{\lambda}{\eta} = const~ Re_{\lambda}^{1/2}  \eqno{(A.4)}
$$
where $const = 15^{1/4}\simeq 2$. Hence
$$
Re_{\lambda}^{eff}\sim \frac{\Gamma}{4\pi \nu}\ln (Re_{\lambda})
\eqno{(A.5)}
$$
The strength $\Gamma$ can be estimated as
$$
\Gamma \sim 2\pi v_{\eta}\eta  \eqno{(A.6)}
$$
where $v_{\eta}=\nu / \eta$ is the velocity scale for the
Kolmogorov (or viscous) space scale $\eta$ \cite{my}. 
Substituting (A.6) into (A.5) we obtain
$$
Re_{\lambda}^{eff}\sim \ln (Re_{\lambda}). \eqno{(A.7)}
$$
Thus, for turbulence processes determined by the vortex
instabilities the relevant dimensionless characteristic is $\ln
(Re_{\lambda})$ rather than $Re_{\lambda}$ (cf Eq. (7) and Fig. 5).

\begin{center}
{\bf Appendix B}
\end{center}
%%%%%%% FIGURE 9 %%%%%%%%%%%%%%%%%%
\begin{figure} \centering
\epsfig{width=.45\textwidth,file=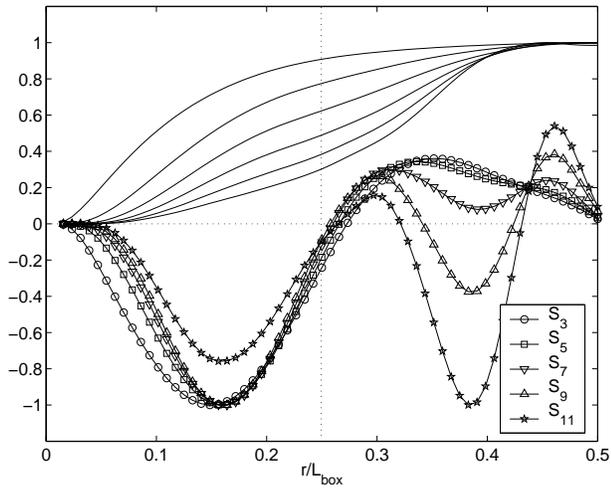} 
\caption{Moments $S_p$ versus $r/L_{box}$ obtained in 
a direct numerical simulation of fluid turbulence on a $2\pi$ cube 
at $128^3$ resolution $Re_\lambda \approx 55$ \cite{d}. Upper curves 
correspond to $p=2,4,6,8,10,12$, the curves with symbols correspond to 
$p=1,3,5,7,9,11$. For $r/L_{box} \geq 0.5$ the picture is 
a mirror reflection of that shown in the figure. }
\end{figure}
%%%%%%%%%%%%%%%%%%%%%%%%%%%%%%%%%%%
In this Appendix we will show that modulation 
of the CMB by the primordial turbulence at the last scattering surface can exhibit 
Gaussian properties much more strong than those of the 'normal' fluid turbulence, 
while preserving the turbulence properties intact. 

We always observe the CMB photons emitted from the 'boundary' (the last scattering surface). 
Therefore, definition of the 'boundary conditions' for the dynamic universe is crucial in 
the case of the CMB radiation. Naturally this is a very 
non-trivial task to define boundary conditions for the universe (horizon). In the numerical 
simulations of the homogeneous turbulence dynamics the periodic boundary conditions are often used. 
The periodic boundary conditions seem to be also a natural suggestion for the universe.

When one considers modulation of the CMB radiation by the Doppler effect 
only the relative difference in velocity between the observer and the source, separated by a distance 
$r$ along the line of sight: $\Delta u_r = u(x+r) - u(x)$, needs to be considered. 
Let us consider the moments of the velocity difference:
$$
S_p (r) = \langle [ u(x)-u( x+r)]^{p} \rangle  \eqno{(B.1)}
$$    
As an example, Fig. 9 shows results of a direct numerical simulation \cite{d}
of fluid turbulence with the periodical boundary conditions
on a $2\pi$ cube (size $L_{box}$) at $128^3$ resolution $Re_\lambda\approx 55$. The curves in the upper 
part of the figure correspond to the even-order moments (B.1), while the curves with symbols 
correspond to the odd-order moments. For $r/L_{box} > 0.5$ the picture is a mirror 
reflection of that shown in the Fig. 9. One can see that the Gaussian conditions: 
$S_p=0$, for the {\it odd}-order moments are satisfied at the 
observer point $r/L_{box} =0.5$. Actually, this is a direct consequence 
of the mirror symmetry provided by the periodic boundary conditions. 
Using the dynamical equations of the baryon-photon fluid it can be shown that 
the even-order moments are also consistent with the Gaussian conditions at the observer point $r/L_{box} =0.5$.


\begin{thebibliography}{99}
\bibitem{p1} F. Sylos Labini, M. Montuori and L. Pietronero, Phys. Rep.,
{\bf 293} 61 (1998)
\bibitem{sbar} K. Subramanian and J.D. Barrow.
Phys. Rev. D, {\bf 58}, 083502 (1998).
\bibitem{dky} R. Durrer, T. Kahniashvili, and A. Yates,
Phys. Rev. D, {\bf 58}, 123004 (1998).
\bibitem{jaf} T.R. Jaffe, A.J. Banday, H.K. Eriksen, K.M.
Gorski, F.K. Hansen, ApJ, {\bf 629}, L1 (2005).
\bibitem{sb1} K.R. Sreenivasan and A. Bershadskii, J.\ Fluid Mech.,
{\bf 54} 477 (2006).
\bibitem{bs1} A. Bershadskii and K.R. Sreenivasan, Phys. Lett. A, {\bf 299},
149 (2002). 
\bibitem{gibson} C.H. Gibson, Appl., Scien. Res., {\bf 72}, 161, (2004).
\bibitem{brand} A. Brandenburg, Astrophys. J., {\bf 550}, 824 (2001).
\bibitem{vc} E. Vishniac and J. Cho, Astrophys. J., {\bf 550}, 752 (2001).
\bibitem{do} A. Dolgov and D. Grasso, Phys. Rev. Lett., 
{\bf 88}, 011301 (2002).
\bibitem{kleo} N. Kleeorin, D. Moss, I. Rogachevskii, and D. Sokoloff, A\&A, {\bf 400}, 9 (2003).
\bibitem{vlc} E. Vishniac, A. Lazarian, and J. Cho, Lect. Notes Phys. {\bf 614}, 376 (2003).
\bibitem{sub} K. Subramanian, A. Shukurov, and N.E.L Haugen, MNRAS {\bf 366}, 1437 (2005).
\bibitem{kah} T. Kahniashvili, astro-ph/0605440 (2006).
\bibitem{tag} M. Tegmark, A. de Oliveira-Costa, A. Hamilton, Phys. Rev. D, {\bf 68}, 123523 (2003)
(http://space.mit.edu/home/tegmark/wmap.html). 
\bibitem{my} Monin A. S. and Yaglom A.M., 1975 Statistical Fluid Mechanics:
Mechanics of Turbulence, V. 2 (The MIT Press, Cambridge).
\bibitem{sa} K.R. Sreenivasan and R.A. Antonia, Annu. Rev. Fluid Mech, {\bf 29},
435 (1997).
\bibitem{b} A. Bershadskii, Phys. Rev. Lett., {\bf 90}, 041101 (2003).
\bibitem{molchan} G. Molchan, private communication.
\bibitem{lg} M.R. Leadbetter and J.D. Gryer, Bull. Amer. Math.
Soc. {\bf 71}, 561 (1965).
\bibitem{pkw} B.R. Pearson, P.-A. Krogstad and W. van de Water,
Phys.\ Fluids, {\bf 14}, 1288 (2002).
\bibitem{sd} K.R. Sreenivasan and B. Dhruva, Prog.\ Theor.\ Phys.\
Suppl., {\bf 130}, 103 (1998).
\bibitem{po} A. Praskovsky and S. Oncley, Fluid Dyn.\ Res.\ {\bf
21}, 331 (1997).
\bibitem{lp} V.S. L'vov and I. Procaccia, Phys. Rev. Lett., {\bf
74}, 2690 (1995).
\bibitem{bt} A. Bershadskii and A. Tsinober, Phys. Lett. A, {\bf 165}, 37 (1992).
\bibitem{cgh} B. Castaing, Y. Gagne and E.J. Hopfinger, Physica D,
{\bf 46}, 177 (1990).
\bibitem{bg} G.I. Barenblatt and N. Goldenfeld,
Phys. Fluids {\bf 7}, 3078 (1995).
\bibitem{ov} J.P. Ostriker and E.T. Vishniac, Astrophys. J,
{\bf 306}, L51 (1986).
\bibitem{batchelor} G.K. Batchelor, An Introduction to Fluid Dynamics 
(Cambridge University Press, Cambridge, 1970).
\bibitem{d} D. Donzis, private communication.

\end{thebibliography}
\end{document}